\documentclass{nsr}

\usepackage{amsmath,graphicx,array}
\usepackage{dcolumn,soul}%

\usepackage{amsthm}
\usepackage[figuresright]{rotating}%
\usepackage{algorithm, algorithmicx, algpseudocode}
\usepackage{listings}%
\usepackage{hyperref}

\makeatletter
\def\uns{\ifmmode\,\else$\,$\fi}%

\makeatother
 
\jvol{XX}
\jnum{X}
\jyear{Year}
\doi{10.1093/nsr/XXXX}
\received{XX XX Year}
\revised{XX XX Year}
\accepted{XX XX Year}

\begin{document}

\dhead{RESEARCH ARTICLE}

\subhead{Life Science}

\title{Infer metabolic velocities from moment differences of molecular weight distributions}

\author{Li Tuobang$^{1,*}$}

\affil{$^1$Li Tuobang is a Chinese student; he thank Mingxun Wang for sharing MassQL. }



\authornote{\textbf{Corresponding authors.} Email: tuobang@biomathematics.org}

\abstract[ABSTRACT]{Metabolic pathways are fundamental maps in biochemistry that detail how molecules are transformed through various reactions. The complexity of metabolic network, where a single compound can play a part in multiple pathways, poses a challenge in inferring metabolic balance changes over time or after different treatments. Isotopic labeling experiment is the standard method to infer metabolic flux, which is currently defined as the flow of a single metabolite through a given pathway over time. However, there is still no way to accurately infer the metabolic balance changes after different treatments in an experiment. This study introduces a different concept: molecular weight distribution, which is the empirical distribution of the molecular weights of all metabolites of interest. By estimating the differences of the location and scale estimates of these distributions, it becomes possible to quantitatively infer the metabolic balance changes even without requiring knowledge of the exact chemical structures of these compounds and their related pathways. This research article provides a mathematical framing for a classic biological concept.

}



\maketitle

\section{Introduction}\label{sec1}

Metabolic pathways consist of enzyme-mediated biochemical reactions that are commonly categorized into two main processes within a living organism: biosynthesis (known as anabolism) and breakdown (known as catabolism) of molecules. It is common to compare the concentration changes of compounds in the same metabolic pathway between two groups of samples, i.e., to assess the up- or down-regulation of a certain pathway. The definitions of up-regulation and down-regulation are actually derived from the principle of chemical equilibrium shifts. For example, the overall equation of the urea cycle can be simplified as $2 \textbf{NH}_3 + \textbf{CO}_2 + 3 \textbf{ATP} + 3 \textbf{H}_2\textbf{O} \rightarrow \textbf{urea} + 2 \textbf{ADP}  + 4 \textbf{Pi} + \textbf{AMP}$. Traditionally, if the concentration of urea, ADP, Pi, or AMP in the experimental group is higher than in the control group, and the concentration of ammonia, carbon dioxide, or ATP is lower in the experimental group compared to the control group, biochemists would say that the urea cycle is up-regulated. This definition stems from the irreversible nature of this cycle and is analogous to the equilibrium shift in chemistry. Since the urea cycle is a synthetic reaction, it is sometimes said that the anabolic process is dominant. Conversely, it is described as down-regulated, and the catabolic process is dominant.

However, this definition is flawed. Even when comparisons are made within the same individuals over time, the change in the amount of certain compounds cannot conclusively determine the direction of the balance shift of a specific pathway, as one compound can be part of several pathways. For example, although urea is a product of the urea cycle, it can also be a product of other metabolic pathways. For instance, arginine, a nitrogen-containing amino acid, can be converted into L-ornithine and urea through the catalysis of L-arginine amidinohydrolase. Additionally, urea serves as the starting material for many metabolic pathways. It can be directly eliminated from the body, converted into carbon dioxide, or synthesized into allophanic acid. This means that if the urea concentration in the experimental group increases, several possibilities could be responsible: the metabolic pathway from arginine to ornithine may be up-regulated, or the downstream pathways may be blocked for some reason in the experimental group.

In practice, it is usually necessary to manually compare the concentration changes of multiple compounds before drawing a conclusion about changes in metabolic balance; however, such conclusions may still be unclear. This article aims to introduce a different approach to quantitatively infer the directions of shifts in metabolic balance for metabolites of interest, without requiring their exact chemical structures and specific pathways. The concept, metabolic velocity, offers a more accessible and biologically explainable framework, with the potential to significantly advance our understanding of metabolic pathways.

\section{Definitions of metabolic velocities}\label{sec2}
Traditionally, a synthesis reaction is defined as process in which two or more simple elements or compounds combine to form a more complex product. For a bimolecular reaction, it is often represented as $\textbf{A}+\textbf{B}\rightarrow \textbf{AB}$. Suppose the molecular weights of A and B are $a$ and $b$ respectively. According to Lavoisier's law of conservation of mass, before the reaction, there are two molecules with an average molecular mass of $\frac{a+b}{2}$, after the reaction, there is only one molecule with a molecular mass of $a+b$. Since $a>0$ and $b>0$, $a+b>\frac{a+b}{2}$.

The above inequality reveals that, for a synthesis reaction, a key hallmark is the increase in average molecular weight. The same principle applies to decomposition reactions. Based on this principle, this article can provide a precise definition of when the anabolic process is dominant and when the catabolic process is dominant.

Suppose, in a biochemical environment, there are $n$ molecules of interest that are known to be interrelated through some chemical reactions. Denote these molecules as $\textbf{M}_1$, $\textbf{M}_2$,$...$, $\textbf{M}_n$. Their molecular weights are $M_1$, $M_2$, $...$, $M_n$, and their molar concentrations are $c_{\textbf{M}_1}$,  $c_{\textbf{M}_2}$, $...$, $c_{\textbf{M}_n}$, in units of molarity. The average molecular weight of these $n$ compounds of interest is given by 
\begin{equation}
Mn=\frac{c_{\textbf{M}_1}M_1+c_{\textbf{M}_2}M_2+...+c_{\textbf{M}_n}M_n}{c_{\textbf{M}_1}+c_{\textbf{M}_2}+...+c_{\textbf{M}_n}}.
\label{eq25}
\end{equation}
In the same study, let the average molecular weight of these $n$ molecules of interest in sample $A$ be denoted as $Mn_{A}$ and that in sample $B$ as $Mn_{B}$. If $Mn_{A}<Mn_{B}$, it is considered that the anabolic process is dominant in sample $B$ compared to sample $A$ with regards to the $n$ molecules of interest. Conversely, if $Mn_{A}>Mn_{B}$, the catabolic process is dominant in sample $B$, meaning that the metabolic balance shifts towards catabolism. This provides a mathematical definition for this classic biological concept.

\begin{figure}
\centering
\includegraphics[width=1\linewidth]{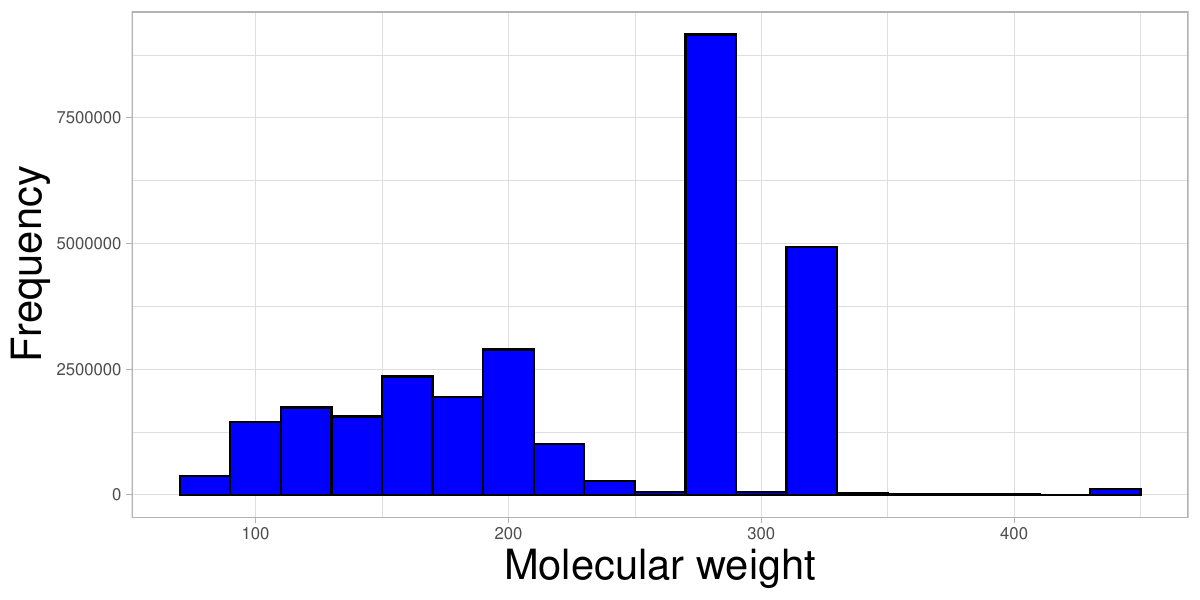}
\caption{The molecular weight distribution of all GC-MS metabolites in human plasma \cite{metabo13080944}
. Arithmetic mean: 233.318; Hodges-Lehmann estimator: 238 \cite{hodges1963estimates}; sample median: 290. Sample standard deviation: 76.277; Bickel-Lehmann spread: 69.598 \cite{bickel2012descriptive}.}
\label{fig1}
\end{figure}

Since the concentration is measured in units of molarity, a molecular weight distribution (MWD) can be formed by replicating the molecular weight of each metabolite of interest, with the replication proportional to the concentration of each metabolite (Figure \ref{fig1}). $Mn$ is essentially the sample mean of the molecular weight distribution (MWD), where the molecular weights are those of the $n$ metabolites of interest. More generally, the location estimate of the MWD for sample $A$ is denoted as $\hat{L}_{n,A}$. The absolute difference between $\hat{L}_{n,A}$ and $\hat{L}_{n,B}$ represents the magnitude of this directional change. This magnitude can be further standardized by dividing it by $\frac{1}{2}(\hat{L}_{n,A}+\hat{L}_{n,B})$. The standardized difference, $\frac{2(\hat{L}_{n,A}-\hat{L}_{n,B})}{(\hat{L}_{n,A}+\hat{L}_{n,B})}$, is called the metabolic velocity of the $n$ molecules of interest from sample $A$ to sample $B$ with respect to location.

\begin{table*}
\centering
\caption{Concentrations of metabolites related to nucleotide metabolism in the lysosome and in the whole cell\label{tab1}}%
\tablefont
  \begin{tabular}{|l|l|l|l|}
  \hline

  compound name       & molecule weight & Whole-cell & Lysosome  \\ \hline
allantoin           & 158.12          & 13.27      & 14.13     \\ \hline
ADP                 & 427.20          & 75.11      & 9.00      \\ \hline
AMP                 & 347.22          & 10.92      & 8.26      \\ \hline
uridine             & 244.20          & 0.88       & 8.10      \\ \hline
guanosine           & 283.24          & 0.25       & 4.19      \\ \hline
inosine             & 268.23          & 0.11       & 2.44      \\ \hline
cytidine            & 243.22          & 0.22       & 2.11      \\ \hline
adenosine           & 267.24          & 0.05       & 1.25      \\ \hline
GMP                 & 363.22          & 3.37       & 0.83      \\ \hline
methylthioadenosine & 297.33          & 1.36       & 0.15     \\ \hline

\end{tabular}

\label{tab:comparison}
\begin{minipage}{1\linewidth}
\footnotesize The unit of molar mass is g/mol. The unit of concentration is $\mu$M.
\end{minipage}
\end{table*}

Then, consider the scale estimate of the MWD for sample $A$, denoted as $\hat{S}_{n,A}$. If $\hat{S}_{n,A}>\hat{S}_{n,B}$, indicating a significant decrease in the scale estimate, the metabolic balance shifts towards centrabolic in sample $B$ compared to sample $A$ for the $n$ molecules of interest. Conversely, sample A is considered duobolic compared to sample B for $n$ molecules of interest. This mathematical approach reveals two new metabolic directions with clear biological significance. If the metabolic direction of a sample of $n$ molecules of interest is centrabolic compared to that of another sample with the same $n$ molecules, it indicates that, for low molecular weight compounds, the related pathways generally shift towards anabolism, while for high molecular weight compounds, the related pathways generally shift towards catabolism. $|\hat{S}_{n,A}-\hat{S}_{n,B}|$ is the magnitude of this change, which can be further standardized by dividing it by $\frac{1}{2}(\hat{S}_{n,A}+\hat{S}_{n,B})$. The standardized difference, $\frac{2(\hat{S}_{n,A}-\hat{S}_{n,B})}{(\hat{S}_{n,A}+\hat{S}_{n,B})}$, is called the metabolic velocity from sample $A$ to $B$ for the $n$ molecules of interest with respect to scale. Analogously, higher-order standardized moments of the MWD for sample A with $n$ molecules of interest can be denoted as $\hat{\mathbf{k}SM}_{n,A}$. However, their biological significance is much weaker. Here, the sample mean and sample standard deviation are used as the location and scale estimators, given that the MWDs are limited to a relatively small range. If the MWD is highly skewed and has a wide range, the Hodges-Lehmann estimator \cite{hodges1963estimates} and Bickel-Lehmann spread \cite{bickel2012descriptive} are recommended as the location and scale estimators. The overall picture of metabolic velocities across different classes is referred to as the velocitome (Table \ref{tab:comparison1}). 


\section{Applications: Targeted Metabolomics}\label{sec1}
Abu-Remaileh et al. determdetermined the concentration of metabolites related to nucleotide metabolism in the lysosome and in the whole cell (Table \ref{tab1}) \cite{science1}. For the whole cell, the sample mean of the molecular weight distribution of these metabolites is 378.89, and the sample standard deviation is 90.66. For the lysosome, the sample mean is 276.65, and the standard deviation is 95.60. The metabolic velocities from the whole cell to the lysosome are -0.05 for location and 0.31 for scale. This indicates that the metabolic balance shifts towards a more catabolic and duobolic state in the lysosome compared to the whole cell. This is consistent with the central role of the lysosome in autophagy \cite{rabanal2018mtorc1}.

\section{Applications: Untargeted Metabolomics}\label{sec2}


In mass spectrometry-based untargeted metabolomics experiments, typically only 10-30\% of the mass spectra can be annotated with specific structures. However, the mass-to-charge ratio (m/z) of each molecule can always be identified. Additionally, compounds within the same chemical classes are generally interrelated. Therefore, besides metabolic pathways, chemical classes can also be used to classify metabolites of interest. As a result, the molecular weight distribution can always be generated without requiring exact structures of the compounds.

The study by Yang et al. compares the plasma metabolome of ordinary convalescent patients with antibodies (CA), convalescents with rapidly faded antibodies (CO), and healthy subjects (H) \cite{Yang2}. For both CA and CO, purine-related metabolism shows a shift towards catabolism and centrabolism compared to the healthy volunteers (Table \ref{tab:comparison1}), which aligns with a previous study indicating that purine metabolism, the hydrolysis of phosphate molecules into nucleosides, is significantly up-regulated after SARS-CoV-2 infection \cite{Xiao_Nie_}. Acylcarnitine-related pathways also exhibit a tendency towards catabolism and centrabolism (Table \ref{tab:comparison1}). This conclusion, which does not require knowledge of individual compounds within the acylcarnitine class, was also emphasized by Yang et al. \cite{Yang2}. It was observed that long-chain acylcarnitines were generally lower in both convalescent groups, while medium-chain acylcarnitines displayed the opposite pattern \cite{Yang2}. For both CA and CO, metabolism related to carbohydrates shifts towards anabolism and centrabolism compared to healthy volunteers (Table \ref{tab:comparison1}). This might be due to elevated glucose levels in COVID-19 patients \cite{salukhov2023impact}. Additionally, pathways related to organoheterocyclic compounds are shown to lean towards centrabolism, while benzenoid-related pathways shift towards anabolism and duobolism.


\begin{table*}
\centering
\caption{Significant velocities of Yang et al.'s UHPLC-MS dataset}
\tablefont
    \begin{tabular}{|l|l|l|l|l|l|l|l|l|}
    \hline
    Compound Class               & Group &   $\bar{x}$ & sd & Comparisons & $\upsilon \bar{x}$ & $\upsilon$sd \\ \hline
Acyl carnitines              & H     & 208.02 & 29.51 & H-CA        & 0.00       & 0.11        \\ \hline
Acyl carnitines              & CO    & 208.20 & 25.70 & H-CO        & 0.00       & 0.14        \\ \hline
Acyl carnitines              & CA    & 207.12 & 26.34 & CA-CO       & -0.01      & 0.02        \\ \hline
Benzenoids                   & H     & 138.96 & 10.10 & H-CA        & -0.01      & -0.33       \\ \hline
Benzenoids                   & CO    & 145.66 & 18.73 & H-CO        & -0.05      & -0.60       \\ \hline
Benzenoids                   & CA    & 140.44 & 14.15 & CA-CO       & -0.04      & -0.28       \\ \hline
Carbohydrates                & H     & 179.40 & 9.70  & H-CA        & 0.00       & 0.11        \\ \hline
Carbohydrates                & CO    & 179.56 & 8.73  & H-CO        & 0.00       & 0.11        \\ \hline
Carbohydrates                & CA    & 179.55 & 8.65  & CA-CO       & 0.00       & -0.01       \\ \hline
Organoheterocyclic compounds & H     & 130.84 & 9.94  & H-CA        & 0.01       & 0.47        \\ \hline
Organoheterocyclic compounds & CO    & 129.98 & 7.03  & H-CO        & 0.01       & 0.34        \\ \hline
Organoheterocyclic compounds & CA    & 129.80 & 6.15  & CA-CO       & 0.00       & -0.13       \\ \hline
Purines                      & H     & 350.53 & 6.59  & H-CA        & 0.00       & 0.09        \\ \hline
Purines                      & CO    & 348.85 & 5.03  & H-CO        & 0.00       & 0.27        \\ \hline
Purines                      & CA    & 349.81 & 6.04  & CA-CO       & 0.00       & 0.18   \\      \hline
      
    \end{tabular}

\label{tab:comparison1}
\begin{minipage}{1\linewidth}
\footnotesize Note: The computations were performed in the same manner as in Table 1, except that the metabolites of interest were not from the entire dataset, but subsets corresponding to compound classes. Only the compound classes having at least one significant change ($\geq$0.1) between groups are listed; others can be found in the SI Dataset S1.
\end{minipage}
\end{table*}

\section{Discussions}\label{sec5}
Since the discovery of zymase by Buchner and Rapp in 1897 \cite{Eduard} and the urea cycle by Krebs and Henseleit in 1932 \cite{Krebs_Henseleit_1932}, a vast body of knowledge on metabolic pathways has accumulated over the last century, especially with the development of analytical techniques such as chromatography, NMR, and mass spectrometry. Metabolomics refers to the large-scale study of small molecules. High-throughput mass spectrometry experiments can collect thousands of mass spectra in just minutes, providing a unique advantage over other analytical methods. The fragmentation pattern of a molecule, or the mass spectrum, can offer valuable structural information about the molecule. However, the annotation of these spectra is typically limited to compounds for which reference spectra are available in libraries or databases \cite{Kind1,Dührkop_Shen_Meusel_Rousu_Böcker_2015,Wang_Carver_Phelan_,Wang_Jarmusch_Vargas0}. Only a small fraction of spectra can be accurately assigned precise chemical structures in nontargeted tandem mass spectrometry studies, which is a prerequisite for pathway analysis \cite{Dorrestein880,Dührkop_Nothias_}. Moreover, many metabolic pathways are still undiscovered or poorly understood, so in practice, often more than half of the metabolites cannot be assigned to any pathways.

Recent developments of in silico methods in class assignment of nontargeted mass spectrometry data can achieve very high prediction performance \cite{Lowry_Isenhour_Justice_McLafferty_Dayringer_Venkataraghavan_1977,Watrous_Roach_Alexandrov_Heath_Yang_Kersten_Voort_Pogliano_Gross,Dührkop_Shen_Meusel_Rousu_Böcker_2015,Nothias_Petras_Schmid_Dührkop_Rainer,Tsugawa_Nakabayashi,Dührkop_Fleischauer_Ludwig_Aksenov_Melnik_Meusel_Dorrestein_Rousu_Böcker_2019b,Aksenov_Laponogov_Zhang_Doran_Belluomo_Veselkov_Bittremieux_Nothias_Nothi,Hoffmann_Nothias_Ludwig_Fleischauer_Gentry_Witting_Dorrestein_Dührkop_Böcker_2021,Petras_Phelan,Morehouse_Clark_,Goldman_Wohlwend_Stražar_Haroush_Xavier_Coley_2023}. The classification of metabolites can be based on chemical characteristics or spectral characteristics \cite{Jarmusch2022}. While this approach can provide replicable information about about changes in metabolites in terms of their chemical properties, it may not directly reflect their interactions within the cell \cite{national2019reproducibility}. Moreover, the total amount of certain classes of metabolites may remain relatively constant within groups, even if the individual compounds within these classes differ.

The classical view of metabolism primarily focuses on individual reactions, resulting in metabolic directions that are mainly considered static, either anabolic or catabolic. This article offers a statistical and holistic perspective on this classic biological concept. The newly defined metabolic velocity has the potential to overcome current limitations and provide fresh insights into biochemistry studies.

\bibliographystyle{nsr}
\bibliography{nsr_sample}

\end{document}